# Anti-bacterial Studies of Silver Nanoparticles

T. Theivasanthi [1] and M. Alagar [2]

[1]Department of Physics, PACR Polytechnic College, Rajapalayam-626108, India.
[2]P.G.Dept. of Physics, Ayya Nadar Janaki Ammal College, Sivakasi-626124, India.

___

**Abstract**

*We discuss about the antibacterial activities of Silver nanoparticles and compare them on both Gram negative and Gram positive bacteria in this investigation. The activities of Silver nanoparticles synthesized by electrolysis method are more in Gram (-) than Gram (+) bacteria. First time, we increase its antibacterial activities by using electrical power while on electrolysis synthesis and it is confirmed from its more antibacterial activities (For Escherichia coli bacteria). We investigate the changes of inner unit cell Lattice constant of Silver nanoparticles prepared in two different methods and its effects on antibacterial activities. We note that slight change of the lattice constant results in the enhancement of its antibacterial activities.*

___

## INTRODUCTION

For centuries, People have used silver for its antibacterial qualities. Nanoparticles usually have better or different qualities than the bulk material of the same element. In the case of silver nanoparticles the antibacterial effect is greatly enhanced and because of their tiny size. Nanoparticles have immense surface area relative to volume. Therefore minuscule amounts of Silver Nanoparticles can lend antimicrobial effects to hundreds of square meters of its host material.

Nanomaterials are the leading requirement of the rapidly developing field of nanomedicine, bionanotechnology. Nanoparticles are being utilized as therapeutic tools in infections, against microbes thus understanding the properties of nanoparticles and their effect on microbes is essential to clinical application. Among noble metal nanoparticles, silver nanoparticles have received considerable attention owing to their attractive physicochemical properties.

Ag-nanoparticles have already been tested in various field of biological science, drug delivery, water treatment and an antibacterial compound against both Gram (+) and Gram (-) bacteria by various researchers. Most of the bacteria have yet developed resistance to antibiotics and in this view in future it is need to develop a substitute for antibiotics. Ag-nanoparticles are attractive as these are non-toxic to human body at low concentration and having broad-spectrum antibacterial nature. Ag nanoparticles inhibit the bacterial growth at very low concentration than antibiotics and as of now no side effects are reported.

**Silver Nanoparticles**
We have synthesized Silver Nanoparticles for these antibacterial activities studies by dissolving Silver Nitrate salt in distilled water and electrolyzed. The Silver Nanoparticles are formed at the cathode and they are removed carefully. XRD analysis of these Nanoparticles has been done and lattice constant calculated.

___

[1] **Corresponding author.**     *E-mail*: theivasanthi@pacrpoly.org

**Changes in Lattice Constant**

Cubic structure of Silver Nanoparticles prepared in electrolysis method is recognized from XRD measurement. Particle size is calculated by using Debey-Scherrer formula.

$$D = \frac{0.9\lambda}{\beta cos\theta} \qquad \text{...........…………….. (1)}$$

Where 'λ' is wave length of X-Ray (0.1541 nm), 'β' is FWHM (full width at half maximum), 'θ' is the diffraction angle and 'D' is particle diameter size. The calculated particle **size is 24 nm**. The value of lattice constant '*a* 'was calculated for (1 1 1) diffraction peak and its calculated value is 4.070 Å. For studying, slight changes in lattice constant of Silver Nanoparticles and its effects on antibacterial activities of Silver Nanoparticles, we have compared this lattice constant '*a* 'value with the Extra cellular synthesis of Silver Nanoparticles by Rajesh.W.Raut et al. [4] and the details are in Table.1.

**Table.1. Comparision of Lattice constant value'*a*' and antibacterial activities of Silver Nanoparticles on Escherichia Coli (Gram Negative bacteria)**

| Silver Nanoparticles synthesis method | Lattice constant value'*a*' ( in Å ) | Zone of Inhibition diameter (in mm) |
|---|---|---|
| Electrolysis method | 4.070 | 12 |
| Extra cellular synthesis method using dried leaves of *pongamia pinnata* (L) pierre | 4.093 | 8 |

**Antibacterial activities evaluation of synthesized Silver nanoparticles in Electrolysis method. (sample no.2)**

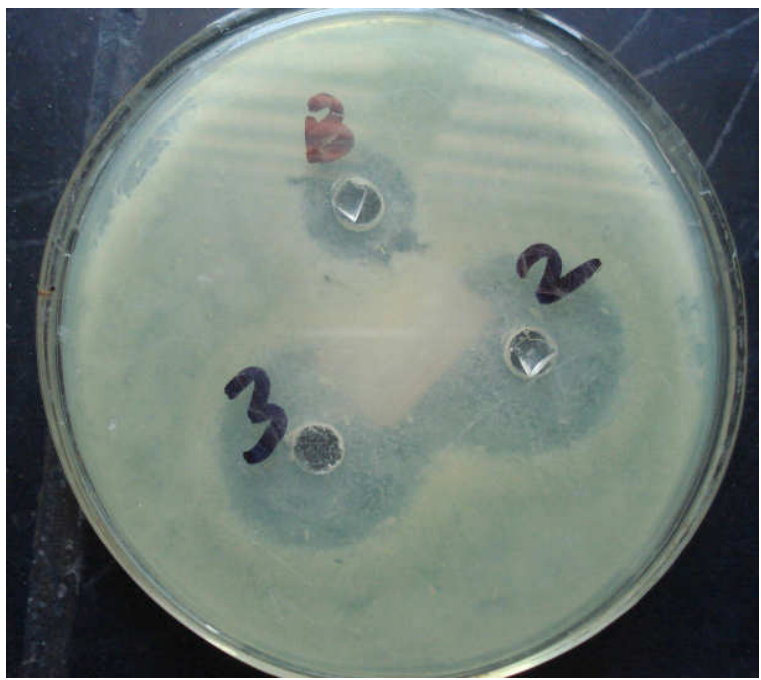

**Figure.1. Zone of inhibition diameter against Escherichia coli bacteria 12 mm**

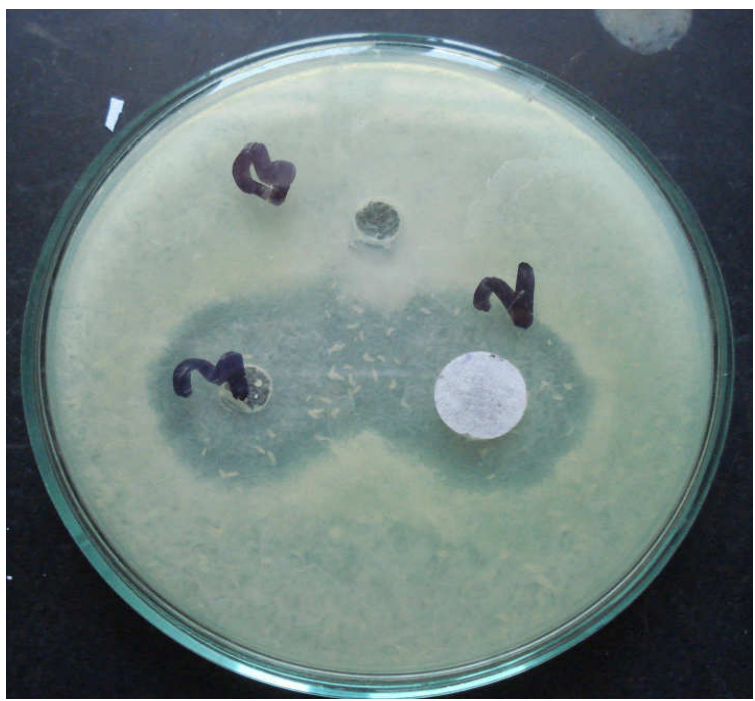

**Figure.2. Zone of inhibition diameter against Bacillus megaterium 6mm**

In order to disclose the effective factors on their antibacterial activity, many studies have already been focused on the powder characteristics, such as specific surface area, particle size and lattice constant by various researchers. Yamamoto et al. have studied the effect of lattice constant on antibacterial activity of ZnO, resulting in the enhancement in antibacterial activity with the slight increase of the lattice constant [2]. However, it is not yet clear what change in antibacterial activity is expected by changing the lattice constant of Silver Nanoparticles. From this comparative study, antibacterial activities of Silver Nanoparticles prepared in Electrolysis method is more on Escherichia Coli (Gram Negative bacteria) than the Silver Nanoparticles prepared Extra cellular synthesis method. We also noted that the slight change of the lattice constant results in the enhancement of antibacterial activities of Silver nanoparticles.

**Silver Nanoparticles & Its Antibacterial Activities**
Silver, a naturally occurring element, is non-toxic, hypoallergenic, does not accumulate in the body to cause harm and is considered safe for the environment. Many manufactured goods like washing machines, air conditioners and refrigerators are using linings of silver nanoparticles for their antimicrobial qualities. Sportswear, toys and baby articles, food storage containers, HEPA filters, laundry detergent etc. are made with silver nanoparticles. The medical field also is using products with silver nanoparticles, such as heart valves & other implants, medical face masks, wound dressings and bandages.

Nanomaterials are the leading in the field of nanomedicine, bionanotechnology and in that respect nanotoxicology research is gaining great importance. Silver exhibits the strong toxicity in various chemical forms to a wide range of microorganism is very well known and silver nanoparticles have recently been shown to be a promising antimicrobial material. Analysis of bacterial growth showed that the toxicity of silver nanospheres is higher than that of gold nanospheres. In addition, no research has discovered any bacteria able to develop immunity to silver as they often do with antibiotics.

Bacteria depend on an enzyme to metabolize oxygen to live. Silver interferes with the effectiveness of the enzyme and disables the uptake of oxygen killing them. This process has the added benefit of not harming humans. A cell wall is present around the outside of the bacterial cell membrane and it is essential to the survival of bacteria. It is made from polysaccharides and peptides named *peptidoglycan*. There are broadly speaking two different types of cell wall in bacteria, called Gram-positive and Gram-negative. The names originate from the reaction of cells to the Gram stain, a test long-employed for the classification of bacterial species. Gram-positive bacteria possess a thick cell wall containing many layers of peptidoglycan. In contrast, Gram-negative bacteria have a relatively thin cell wall consisting of a few layers of peptidoglycan.

Antibacterial activity of silver nanoparticles synthesized by Electrolysis was evaluated by using standard zone of inhibition (ZOI) microbiology assay. The sample silver nanoparticles prepared in electrolysis method showed diameter of inhibition zone against E.Coli 12 mm & B.megaterium 6mm (The results in Figures 1 and 2). Siddhartha Shrivastava et al [5] concluded that the antibacterial effect of silver nanoparticles was more pronounced against gram (-) than gram (+) bacteria. A comparative analysis was done on the basis of this concept.

**Table.2. Comparision of activities of Silver Nanoparticles on Gram (-) and Gram (+) bacteria**

| Silver Nanoparticles synthesis method | Name of Bacteria | Variety of Bacteria | Inhibition Zone diameter (in mm) |
|---|---|---|---|
| Electrolysis method | Escherichia coli | Gram (-) | 12 |
| | Bacillus megaterium | Gram (+) | 6 |
| Extra cellular synthesis method using dried leaves of *pongamia pinnata* (L) pierre | Escherichia coli | Gram (-) | 8 |
| | Pseudomonas aeruginosa | Gram (-) | 10 |
| | Klebsiella pneumoniae | Gram (-) | 12 |
| | Staphylococcus aureus | Gram (+) | 12 |

**Enhancement of Antibacterial Activities**
People have used silver for its antibacterial qualities for many centuries. However, Silver Nanoparticles have showed antibacterial activities more than silver. In addition to this various researchers have tried to enhance the antibacterial actions of Silver nanoparticles adopting various methods i.e. using capping agents while on synthesis, using a combination of light energy with nanoparticles, using a combination of ultrasound wave with nanoparticles, using a combination of electric field with nanoparticles etc.

Dhermendra K. Tiwari and J. Behari reported that the silver nanoparticles treated with short time exposure with ultrasound show increased antibacterial effect but this time was not enough to kill the bacterial cells with ultrasound only [1]. It indicated that synergistic effect of Ultrasound & Silver nanoparticles. The ultrasound facilitates the entry of Ag-nanoparticles inside the cells and the antibacterial effect was enhanced with same concentration of nanoparticles in presence of ultrasound waves. The Biocidal effect was more pronounced when compared to the actions of Silver nanoparticles alone.

Omid Akhavan and Elham Ghaderi investigated the effect of an electric field on the antibacterial activity silver nanorods against *E. coli* bacteria [3]. It was found that the grown silver nanorods show strong and fast antibacterial activity. Applying an electric field in the direction of the nanorods (without any electrical connection between the nanorods and the capacitor plates producing the electric field) promoted their antibacterial activity. This indicated that the antibacterial activity of silver nanorods can be enhanced by applying an electric field.

In view of the above, we have tried to increase antibacterial activities of Silver Nanoparticles for which we have made an attempt using electrical power while on synthesizing of Silver nanoparticles. We have synthesized Silver nanoparticles in Electrolysis by using electrical power. Silver nanoparticles synthesized in this method have showed more antibacterial activities (For E.Coli bacteria) than Silver nanoparticles synthesized in Extra cellular synthesis method and the details are in Table.2.

## RESULTS AND DISCUSSION

Lattice constant of Silver nanoparticles synthesized in Electrolysis method and Extra cellular synthesis method was calculated and compared. We noted that a slight change in lattice constant of Silver nanoparticles was showing more antibacterial activities. We also noted that using electrical power while on synthesizing of Silver nanoparticles is increasing its antibacterial activities. Actions of Silver nanoparticles synthesized in both above method was also compared against both Gram (-) and Gram (+) bacteria. Antibacterial activities of Silver nanoparticles synthesized in Electrolysis method only were more in Gram (-) than Gram (+) bacteria which agreed with the Silver nanoparticles research of Siddhartha Shrivastava et al [5] but not in the case of another method.

## CONCLUSION

We have come for conclusion that Silver nanoparticles synthesized in Electrolysis method are showing antibacterial activities against both Gram (-) and Gram (+) bacteria but the activities more in Gram (-) than Gram (+) bacteria. A slight change in lattice constant is enhancing its antibacterial activities. Silver nanoparticles synthesized in Electrolysis method are showing more antibacterial activities (For E.Coli bacteria) than Silver nanoparticles synthesized in Extra cellular synthesis method. Using electrical power while on synthesizing of Silver nanoparticles is increasing its antibacterial activities.

**Acknowledgements**
The authors express immense thanks to **Dr.M.Palanivelu**, *Principal of Arulmigu Kalasalingam College of Pharmacy (Kalasalingam University, Krishnankoil, India)*, staff & management of *PACR Polytechnic College*, Rajapalayam, India and *Ayya Nadar Janaki Ammal College*, Sivakasi, India for their valuable suggestions, assistances and encouragements during this work.

## REFERENCES

[1]. Dhermendra K. Tiwari and J. Behari, *"Biocidal Nature of Combined Treatment of Ag-nanoparticle and Ultrasonic Irradiation in Escherichia coli dh5"* Advances in Biological Research. 2009, 3 (3-4), 89-95.

[2]. Osamu Yamamoto, Miyako Komatso, Jun Sawai, and Zenbe-E-Nakagawa, *"Effect Lattice constant of Zinc Oxide on antibacterial characteristics"* Journal Materials science: Materials in Medicine. 2004, 15, 847-851.

[3]. Omid Akhavan and Elham Ghaderi, *"Enhancement of antibacterial properties of Ag nanorods by electric field"* Sci.Technol.Adv.Mater. 2009, 10, 015003 (5pp). doi: 10.1088/1468-6996/10/1/015003.

[4]. Rajesh W. Raut, Niranjan S. Kolekar, Jaya R. Lakkakula, Vijay D. Mendhulkar and Sahebrao B. Kashid, *"Extracellular synthesis of silver nanoparticles using dried leaves of pongamia pinnata (L) pierre",* Nano-Micro Lett. 2010, 2, 106. doi: 10.5101/nml.v2i2.p106-113.

[5]. Siddhartha Shrivastava, Tanmay Bera, Arnab Roy, Gajendra Singh, P Ramachandrarao and Debabrata Dash, *"Characterization of enhanced antibacterial effects of novel silver Nanoparticles",* Nanotechnology. 2007, 18, 225103 (9pp). doi: 10.1088/0957-4484/18/22/225103.